# Making the Guided Wave Propagate Unidirectional in Periodic Dielectric Waveguide

Qingbo Li, [1, 2] Hui Ma [1], and Rui-Xin Wu [1,*]

[1] School of Electronic Sciences and Engineering, Nanjing University, Nanjing 210093, China
[2] School of Physics and Electronic Electrical Engineering, Huaiyin Normal University, Huaian 223300, China

Unidirectional waves are the waves propagating only in one direction and prohibited in opposite direction. This kind of waves can be surface waves realized by topological edge state or bulk waves in the media with time reversal and space inversion symmetry broken. Here, we introduce unidirectional waveguide modes in the periodic dielectric chain comprised of gyromagnetic rods. In this modes, the waves in the waveguide propagates unidirectional, but can be converted into bi-directional waves by changing the arrangement of rods. We show unidirectional guided waves result from the time-mirror symmetry broken of the chain. Experiments confirm the theoretical results of waveguide modes. A new vortex wave generator based on the unidirectional guided wave is demonstrated theoretically. Our work opens a new avenue for manipulating the electromagnetic wave in a simple and flexible way, and provides a new possibility for novel devices.

# 1. Introduction

In conventional materials, light always propagates forward and backward at the same time because of the time reversal (T) symmetry or space inversion (I) symmetry of Maxwell equation [1], the master equation for lights propagating in medium. To make the light propagation only in one direction but prohibited in opposite direction, the materials of the light undergoing are required with both T- and I-symmetry broken [2]. In this circumstance, unidirectional bulk waves may appear. The realization of those waves is proposed in man-made materials, such as gyrotropic multilayers [3], and parity-time symmetry materials [4, 5] and magnetic PCs [6]. By design the configuration of the man-made materials, the unidirectional bulk wave can route even in a desired routine [7]. Another way to realize the one-way wave is the topological states in photonic crystals or metamateials [8-10]. That states are achieved through careful design of wavevector space topologies, which allow light to flow unidirectional at the interface of the materials and to go around large imperfections without back-reflection [11-13]. The one-way surface waves are first experimentally demonstrated in the microwave regime using gyromagnetic PCs [11], where the one-way surface waves reside at the interface of a photonic topological insulator and its non-topological environment. Later, other ways to achieve one-way surface waves in non-magnetic photonic topological insulators are proposed, such as helical structures [14, 15], efficient time-modulated systems [16, 17], ring resonators [18], photonic analogues of the spin and valley Hall effects [19-23].

Now, we will introduce unidirectional waveguide modes in periodic dielectric waveguides (PDWG). We show that waveguide is a one-way waveguide when its time-mirror symmetry is broken in addition to T- and I-symmetry broken. Making the PDWG with T-shaped ferrite rods, we illustrate the waveguide can support both unidirectional or bidirectional waves depends on the waveguide configuration. It is found that one-way waveguide modes may associated with the asymmetrical bulk states of the corresponding 2D magnetic PC. Further, the unidirectional guided wave is observed experimentally in magnetic PDWG composed by T-shaped rods. The one-way waveguide modes provide a new method to control the electromagnetic wave propagation and to create novel devices for abnormal applications. It also makes the one-way waveguide fabrication much easier and flexible.

## 2. One-way guided waves in periodic ferrite chain

### 2.1 Waves in T-shaped periodic ferrite chain

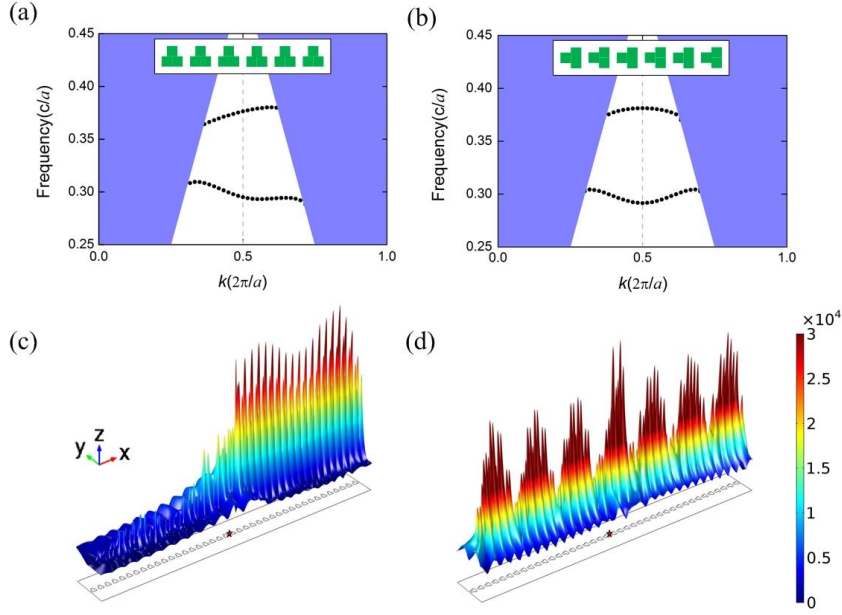

FIG. 1. Dispersion diagram and wave propagation in magnetic periodic dielectric waveguide made of T-shaped ferrite YIG rods. (a) and (b) are the dispersion curves of the waveguide at different arrangement of T-shaped rods, as displayed in the inset of the panels. The shaded blue area is the light cone in which the guided waves will radiate into the air. (c) and (d) are the three-dimensional view of the magnitude distribution of electric field around the waveguide corresponding to the T-shaped rod arrangement in the insets of panels (a) and (b). The results are obtained under the conditions of bias DC magnetic field $H_0$=600 Oe, period $a$=12mm and feature length of T-shaped rod $l$=4 mm (refer to the inset of Fig. 3).

Considering a periodic dielectric chain made of inverted T-shaped ferrite rods in air background as displayed in the inset of Fig. 1(a). The rods are lined along $x$-direction with period $a$. Suppose the ferrite rods are biased by a DC magnetic field along rods' axes that is defined as $z$-axis. In this case, the permeability of the rods is a tensor. At fully magnetized state, the tensor is written as[12]:

$$\ddot{\mu}_r = \begin{pmatrix} \mu & j\kappa & 0 \\ -j\kappa & \mu & 0 \\ 0 & 0 & 1 \end{pmatrix}.$$

The elements of the tensor are, respectively,

$$\mu = 1 + \frac{\omega_m(\omega_0 + j\alpha\omega)}{(\omega_0 + j\alpha\omega)^2 - \omega^2}, \quad \kappa = \frac{\omega\omega_m}{(\omega_0 + j\alpha\omega)^2 - \omega^2}.$$

where, $\omega_0 = 2\pi\gamma H_0$ and $\omega_m = 4\pi\gamma M_s$ are the processing frequency and the characteristic frequency of the magnetic rods, respectively; $\gamma=2.8$ MHz/Oe is the gyromagnetic ratio, $H_0$ local magnetic field and $\alpha$ magnetic damping coefficient of the rods. $\omega$ is the operating frequency.

We use ferrite yttrium-iron-garnet (YIG) to make the rods. The material parameters of YIG are the saturation magnetization $4\pi M_s$ =1884 Oe, and the relative permittivity $\varepsilon_r$ =15.26, respectively. Figure 1(a) plots the dispersion diagram of the YIG chain when period *a*=12mm and bias field $H_0$=600 Oe. We see two separated energy bands are below the light cone shaded in blue. Since the bands is below the light corn, the corresponding electromagnetic modes will be localize modes or waveguide modes for which the electromagnetic waves will be concentrated in the chain and evanescent out of the chain. Thus, the chain resembles a waveguide, which is known as periodic dielectric waveguide [24]. One special in Fig. 1(a) is the dispersion curves of the waveguide modes are monotonous that indicates the group velocity of the guided waves is along one direction, i.e. the flow of EM energy is locked in one direction in the waveguide. To illustrate this one-way propagating characteristic, Fig. 1(c) plots electric field distribution around the PDWG at normalized frequency 0.376 (c/*a*). The simulation result shows the excited wave only goes along the right-hand side and fades sharply in the opposite direction and both sides out of the chain, embodying a good self-guided unidirectional wave propagation characteristic. This wave feature is quite different from the conversional PDWG where the waves propagate bidirectional and the dispersion curve is symmetric about *k*=0 or *k*=0.5(π/*a*) owing to the translation symmetry of the PDWG [25].

In addition to the one-way waves, bidirectional waves can appear in the waveguide, which depends on the arrangement of the rods in the chain. For example, Fig. 1(b) plots the dispersion curve of the waveguide when we rotate the T-shaped rods 90° with respect to the rod arrangement in Fig. 1(a). In this case, the dispersion curve of the waveguide modes is symmetrical. As a result, the wave propagates bidirectional in the chain as displayed in Fig. 1(d). That the feature of wave propagation characteristics can be tuned by the configuration of the waveguide provides a simple way to control the wave propagating.

**2.2 Differences of one-way waveguide modes to other types of one-way modes**

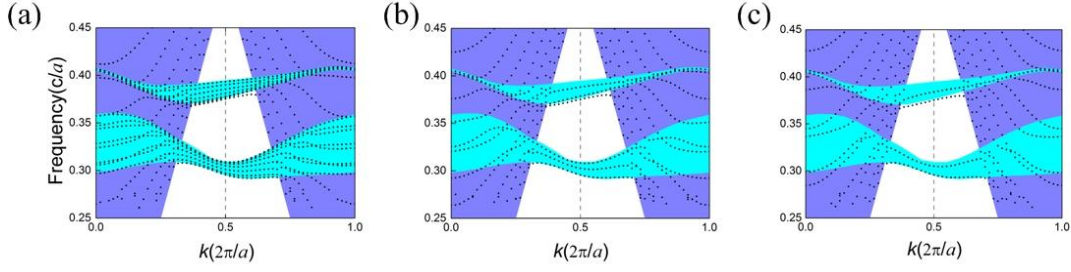

FIG. 2. Band diagram of full and slab of magnetic PC with T-shaped rod in the configuration in Fig. 1(a). In each panels the light green region represents the bands for full magnetic PC and blue region represents the light cone. The dot lines are the dispersion curves for the slabs of PC. For panels (a)-(c), the numbers of rod chains in the slab are 8, 3 and 2, respectively. Out of the light cone, the presence of slab edges does not create any new energy bands, and the dispersion curves are asymmetrical for both full PC and PC slabs. The asymmetrical dispersion is more evident in panel (c).

The unidirectional waveguide modes of the magnetic PDWG in above are different from that of one-way edge states, though they all show unidirectional wave characteristics [8, 9]. To clarify this distinction, we calculate the projected band diagram of full magnetic PC and compared with that of its slabs using supercell technique. The magnetic PDWG is the single chain limit of the slabs. Figure 2(a) shows the results with the rod arrangement corresponding to Fig. 1(a). The shaded region in light green in Fig. 2(a) represents the bands of full PC and the dot lines are the bands of the PC slab with 8 rod chains. We see the presence of slab edges does not introduce any new band below the light line. All the bands are within the region of the full magnetic PC. This result suggests the one-way waveguide modes of magnetic PDWG are not related to the edge states of the PC.

We notice the band structure of full magnetic PC is asymmetric and the asymmetry is more evident outside of the light cone. This feature is maintained in the PC slabs. By decreasing the number of chains in the slab, the number of energy bands below the light line decreases. The results are plotted in the dot lines in panels Figs. 2(a)-(c) where the slabs are with 8, 3, and 2 rod chains, respectively. We see the number of bands is the same as number of rod chain in the slabs. At the single chain limitation, the slab has one band below the light line as displayed in Fig. 1(a). Though the frequency span of waveguide modes changes as number of chain in the slab decreases, it is always within the frequency range of the energy bands of full magnetic PC. This further suggests the unidirectional waveguide modes may associated with the non-reciprocal bulk states of the magnetic PC [3, 6], but it differs from the bulk states since the multiple chains may result in bidirectional waves in the waveguide as can be seen from the band diagram in Fig. 2(a).

**2.3 One-way waveguide mode and the symmetry of T-shaped magnetic rods**

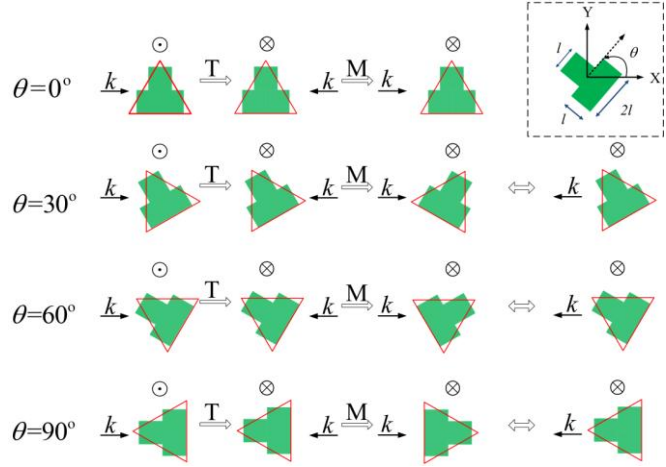

FIG. 3. Schematic diagram of time and mirror (in *x*-direction) operations on the T-shaped rod. The combined operations do not change the direction of wave vector *k* at rotation angle $\theta$=0 and 90 degrees, making the guided waves propagating unidirectional. But the operations do change the direction of wave vector *k* at angle 30 and 60 degrees.

The unidirectional or bidirectional waves in the magnetic PDWG can be explained by the symmetry of the rods. It is known that if the symmetry group G of a periodic structure includes a symmetry operation $\hat{g}$ that changes the sign of the Bloch vector $\vec{k}$, then the spectrum $\omega(\vec{k})$ of Bloch eigen-modes supported by the system will be reciprocal for this particular direction of $\vec{k}$ [26]. For magnetic systems, one has to be sure that the magnetic symmetry group G of the periodic structure does not include any operations that change the sign of the wave vector $\vec{k}$, that is $\hat{g}\vec{k} \neq -\vec{k}$ for any g∈G [26]. In our case, in addition to the time and inversion symmetry broken, it is needed to further break any other symmetry operations (including the combination of the symmetry operations) that map *x*-> -*x*. Since the T-shaped magnetic rods have certain mirror symmetry in *x*- and *y*-direction, the symmetry can affect the wave propagation in the waveguide. Figure 3 schematically displays the results of time and mirror (M) operations (in *x*-direction) on the T-shaped rod under their different arrangement. The arrangement of the rod is characterized by the rotation angle $\theta$ displayed in the upper right panel. When T-operator is applied on the rod, the bias magnetic field and wave vector reverse their direction. The next mirror operation may change or not change the direction of wave vector depending on the rotation status of T-shaped rod. At the rotation angle $\theta=0°$, the figure shows the operations do not change the $k$-direction, but at $\theta=90°$ the $k$-direction changes its direction. Thus, the guided wave is unidirectional in first case but bidirectional in the second case, as has been displayed in Figs. 1(c) and 1(d). The same results are obtained for the mirror operation about $y$-axis, which is not depicted in the figure.

The symmetry analysis is also applied to the rod arrangement at other rotation angles. The rotation symmetry of the T-shaped rod is very close to that of triangle rod that is illustrated in red triangle in the figure. Therefore, we will further consider other two special cases at angles $\theta=30°$ and $60°$. At $\theta=30°$, Fig. 3 indicates the guided waves will be bidirectional; the symmetry operations of T and M change the direction of wave vector. However, at $\theta=60°$, the symmetry operations do not change the k-direction, therefore the waves will be unidirectional, However, the propagation direction will be opposite to the case of $\theta=0°$ since the T-shaped rod is approximately a mirror of the case $\theta=0°$. For the other cases, the guided waves in magnetic PDWG will be unidirectional.

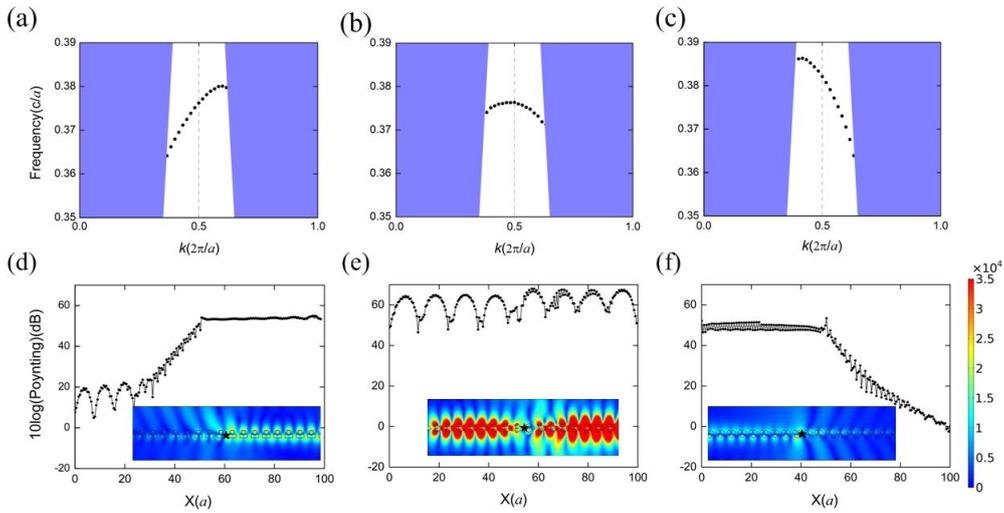

FIG. 4. Dispersion diagram and electric field distribution in the waveguide. Panels (a)-(c) are the dispersion curves of the waveguide when the T-shaped rod takes the rotation angle $\theta=0°$, $30°$ and $60°$, respectively. The shaded region is the light cone. The dispersion curve in (b) is basically symmetrical, while the curves in (a) and (c) are monotone but with opposite sign of curve slop. Panels (d)-(f) are the electric field distribution along the axis of the waveguide at the frequency at $0.372(c/a)$, which are corresponding to the rotation angles of panels (a)-(c). The inset of each panel is the field distribution near the waveguide, where the black star marks the position of source. The field is mainly localized along the rods.

To illustrate the results of symmetry analysis, in Figs. 4(a)-(c), we plot the dispersion curves of the waveguide in the normalize frequency range 0.35 to 0.39($c/a$) at the angles $\theta$=0, 30 and 60 degrees. At the angle $\theta$=30° the dispersion curve is basically symmetry to the axis of $k$=0.5. Therefore, the guided waves will be basically bidirectional in the waveguide. In contrast, at $\theta$=60° the dispersion curves are asymmetry. However, the slop of the curve, the group velocity of the guided wave is in opposite sign of the case of $\theta$=0°(see Fig. 4(a)) indicating the guided waves propagating unidirectional but in opposite directions.

We verified the wave propagation characteristics by full wave simulations. Figures 4(d)-(f) show the magnitude of power flow along the axis of the waveguide. The insets display the electric field distribution near the waveguide at the normalized frequency 0.372($c/a$). In panels d and f, we see the power flow maintains constant in one direction but declines rapidly in opposite direction. The corresponding field distribution in the inset intuitively shows the field is mainly localized in waveguide and decay out of it. For these two cases, however, the directions of power flow are opposite. In contrast, in panel e the guided wave is bi-directional; the waves propagate in both left and right directions.

**2.4 Experimental verification**

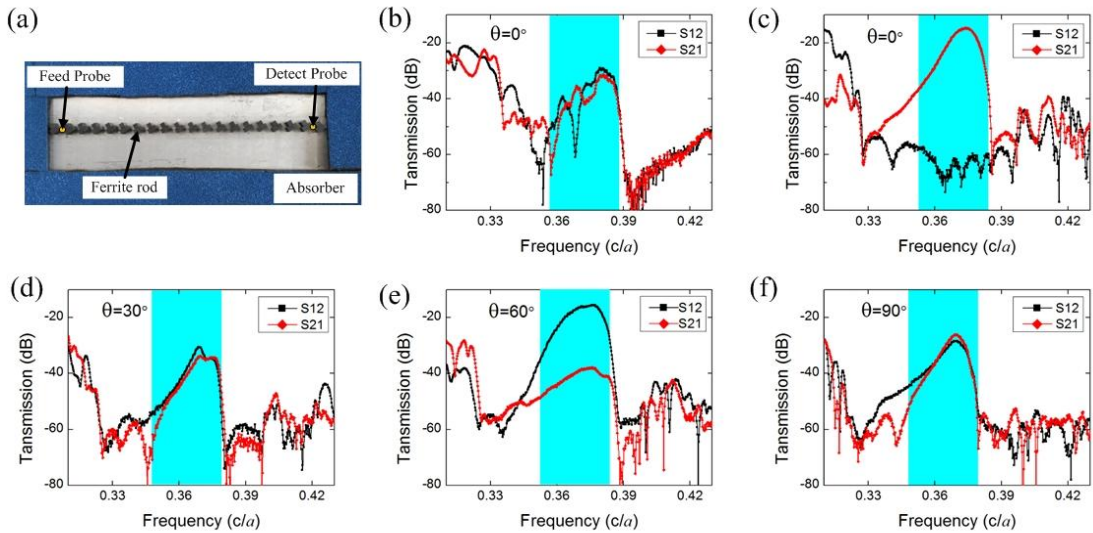

FIG. 5. Experimental results of one-way waveguide. (a) Image of the T-shaped rod chain sample in experiments. (b) Transmission under zero bias magnetic field. The forward and backward transmissions, S21 and S12, are the same. (c)-(f) are the transmissions for the waveguide with different arrangement of T-shaped rods under bias magnetic field $H_0$=600 Oe. Panels (c)-(f) correspond to the rotation angel $\theta$=0, 30, 60 and 90 degrees, respectively. At angles 30 and 90 degree the S21 and S12 are the same, but at 0 and 60 degrees the S21 and S12 have huge difference, showing a nonreciprocal transmission. For the later, the high transmission changes from S21 to S12, indicating the reverse of one-way wave direction.

Experimentally, we fabricated T-shaped magnetic PDWG by ferrite material YIG. Figure 5(a) shows the PDWG sample in experimental setup. The period of the rods is $a$=12mm. In experiments, the sample was put in a parallel-plate waveguide with the same height of the rods and enclosed with absorbers. One probe was used to excite the guided waves, which was polarized perpendicular to the parallel plates. Another probe which was far away from the exiting probe was used to detect the field of the guided wave.

Figures 5(b)-(f) plot the results of transmission measurements. We first measured the waveguide when the T-shaped rods at rotation angle $\theta$=0°. Figure 5(b) plots the transmission results at no bias magnetic field. The forward and backward transmission, S21 and S12, are the same, indicating the waves in the waveguide are bi-directional. When the bias magnetic field $H_0$=600 Oe was applied, however, the forward and backward transmissions show a huge difference. As displayed in Fig. 5(c), the transmission S21 is much greater than S12 in the frequency range from 0.3528($c/a$) to 0.3534($c/a$). At 0.3728 ($c/a$), the distinction between S21 and S12 is over 50 dB, showing good one-way wave characteristics [12]. Figures 5(c)-(f) plot the transmission when T-shaped rods take different arrangements. We see the forward and backward transmissions change a lot in shaded frequency range. When angle $\theta$ takes 30°, as displayed in Fig. 5(d), the guided wave is no long propagating unidirectional but bidirectional. The forward transmission is almost the same as backward transmission. This phenomenon is also observed in the case of $\theta$=90°, which is predicted in Fig. 1(d). In these two cases the time and mirror symmetry operations on T-shaped rods change the direction of wave vector of the guided waves. At angle $\theta$=60°, the guided wave becomes unidirectional, but the transmission S12 is now much greater than S21 as shown in Fig. 5(e). The result indicates the unidirectional waves propagating in the opposite direction of the case $\theta$=0°. All the experimental results are in good agreements with the results of theoretical calculation and the symmetry analysis.

### 3. Potential Applications

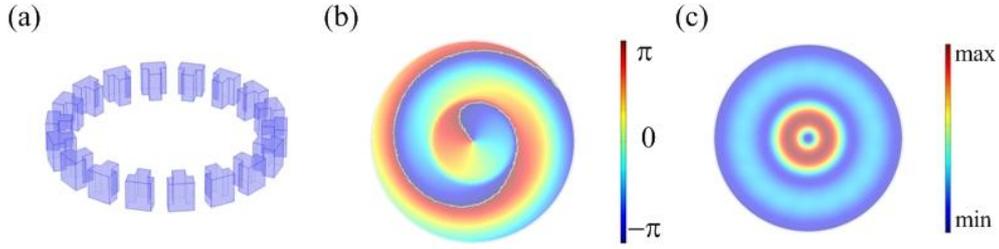

FIG. 6. Prototype of vortex wave generator. (a) Schematic diagram of ring cavity based T-shaped magnetic periodic dielectric waveguide. (b) and (c) are the phase and intensity of the electric field of excited vortex wave above the ring cavity. Simulation results of the spiral phase distribution shows the topological charge equals to one. The intensity distribution displays a doughnut shape, the typical vortex wave profile.

Our one-way waveguide mode can be used to design devices for controlling the EM wave motion. For example, the one-way waveguide can be used as an isolator when the waveguide works in unidirectional waveguide mode, or switcher or diode when waveguide can dynamically switches between the unidirectional mode and bi-directional mode.

Here, we show the waveguide modes can be applied to generate vortex beam. The vortex beam with helical phase front carries orbital angular momentum (OAM) is considered to be an important technology in the significant fields, such as quantum communication [27], super resolution imaging [28], and spinning object detection [29]. OAM can be created by such as ring cavities, however the clockwise and counterclockwise eigen-modes would inevitably cause the corresponding waves travel in opposite directions. Thus the two waves interfere and may cancel each other. One strategy to overcome the cancellation is to use the unidirectional power flow inside the cavity [30, 31]. Following this line, we bend T-shaped rods chain into a circular ring as illustrated in Fig.6 (a). At designated frequency, the ring waveguide functions as a ring cavity in which unidirectional power circulates. Due to the finite Q-factor of the ring cavity, the coupling between the cavity and out of it will generate vortex beam outside of the ring cavity. Figs. 6(b)-(c) show the phase and amplitude of first order vortex beam by simulations. In the simulation, ring cavity with a diameter of 68.7mm has a resonant frequency 10.2GHz. The figures show a phase distribution in a spiral shape and the intensity profile of doughnut shape, which are typical vortex beam characteristics. The continuous phase variation is ranging from $-\pi$ to $\pi$ along the azimuthal direction. The central blue region originates from the phase singularity at the center of the OAM vortex radiation. Our results provide a simple way of generating optical vortex using textual structure. By optimizing the design, this prototype should also be considered at telecommunication wavelengths though it is YIG material [32]. Such flexible design could open up truly large-scale applications opportunities.

**4. Conclusion**

In conclusion, we have studied non-reciprocal wave propagation in one-dimensional periodic dielectric waveguide comprised of gyromagnetic T-shaped rods. We show the guided waves in the PDWG can be unidirectional or bi-directional depended on the arrangement of magnetic rods. The time-mirror symmetry of the waveguide should be response for the different types of wave propagation. The unidirectional guided waves caused by time-mirror symmetry broken are different from one-way waves of topological edge states and one-way bulk states. The transmission experiments further confirm the unidirectional guided waves and the transition between the unidirectional and bi-directional waves by changing the rods arrangement. Featuring the simple and flexibility of our unidirectional waveguide, a possible application of unidirectional waveguide modes, the vortex beam generator, has theoretically demonstrated. The magnetic PDWG in this work provide a new way to manipulate the wave propagation, which may develop new technologies to realize non-reciprocal devices or brand new devices, and can be extended to other frequency range as long as gyromagnetic materials are used.

**Acknowledgements.** This work is supported by the National Natural Science Foundation of China (NSFC) (61771237, 61701187, and 61671232); R-X. W. thanks the partial support from Priority Academic Program Development (PAPD) of Jiangsu Higher Education Institutions.

*Corresponding author**.** rxwu@nju.edu.cn

**References**

[1] L. Landau, E. Lifshitz, L. Pitaevskii, *Electrodynamics of continuous media* (Pergamon, New York, 1984).
[2] A. Figotin and I. Vitebsky, Phys. Rev. E **63**, 066609(2001).
[3] Z. Yu, Z. Wang, and S. Fan, Appl. Phys. Lett. **90**, 121133(2007).
[4] L. Feng, M.Ayache, J. Huang, Y. L. Xu, M. H. Lu, Y. F. Chen, Y. Fainman, A. Scherer, Science **333(6043)**, 729–733(2011).
[5] L. Feng, Y. L. Xu, W. S. Fegadolli, M.H. Lu, J. B. E. Oliveira, V. R. Almeida, Y. F. Chen, A. Scherer, Nat. Materials **12(2)**, 108–113(2013).
[6] C. He, M. H. Lu, X. Heng, L. Feng, and Y. F. Chen, Phys. Rev. B **83**, 075117(2011).
[7] Q. B. Li, Z. Li, and R. X. Wu, Appl. Phys. Lett. **107**, 241907(2015).
[8] T. Ozawa, H. M. Price, A. Amo, N. Goldman, M. Hafezi *et al.* arXiv preprint arXiv:1802.04173(2018).
[9] L. Lu, J. D. Joannopoulos, and M. Soljacic, Nat. Photonics **8**, 821(2014).
[10] D. L. Sounas, C. Caloz, and A. Alù, Nat. Commun. **4**, 2407(2013).
[11] Z. Wang, Y. Chong, J. D. Joannopoulos, and M. Soljacic, Nature (London) **461**,772–775(2009).

[12]Y. Poo, R. X. Wu , Z. F. Lin, Y. Yang, and C. T. Chan, Phys. Rev. Lett. **106**, 093903(2011).
[13]J. Lian, J. X. Fu, L. Gan, and Z. Y. Li, Phys. Rev. B **85**, 125108(2012).
[14]A. Christofi, N. Stefanou, Phys. Rev. B **87**, 115125(2013).
[15]R. Fleury, D. L. Sounas, C. F. Sieck, M. R. Haberman, and A. Alù,Science **343**, 516–519(2014).
[16]Q. Xu, B. Schmidt, S. Pradhan, and M. Lipson, Nature (London) **435**, 325–327(2005).
[17]C. T. Phare, Y. H. Daniel Lee, J. Cardenas, and M. Lipson, Nat. Photonics **9**, 511–514(2015).
[18]Z. Yu and S. Fan, Nat. Photonics **3(2)**, 91–94(2009).
[19]X. Wu, Y. Meng, J. Tian,Y. Huang, H. Xiang, D. Han and W. Wen, Nat. Commun. **8**, 1304(2017).
[20]F. Gao, H. Xue, Z. Yang, K. Lai, Y. Yu, X. Lin, Y. Chong, G. Shvets, B. Zhang, Nat. Physics **14**, 140–144(2018).
[21]Y. Kang, X. Ni, X. Cheng, A. B. Khanikaev, A. Z. Genack, Nat. Commun. **9**, 3029(2018).
[22]M. Yan, J. Lu, F. Li, W. Deng,X. Huang,J. Ma,Z. Liu, Nat. Materials **17**, 993–998(2018).
[23]M. I. Shalaev, W. Walasik, A.Tsukernik, Y. Xu and N. M. Litchinitser, Nat. Nanotechnology **14**, 31–34 (2019).
[24]S. Zeng, Y. Zhang, and B. Li, Opt. Express **17(1)**, 65–378(2009).
[25]P. G. Luan and K. D. Chang, Opt. Express **14(8)**, 3263–3272(2006).
[26]I. Vitebsky, J. Edelkind, E. N. Bogachek, A. G. Scherbakov, and U. Landman, Phys. Rev. B **55**, 12566(1997).
[27]A. C. Peacock, M. J. Steel, Science **351(6278)**, 1152–1153(2016).
[28]P. C. Maurer, J. R. Maze, P. L. Stanwix, L. Jiang *et al*. Nat. Phys. **6**, 912–918(2010).
[29]M. P. J. Lavery, F. C. Speirits, S. M. Barnett and M. J. Padgett, Science **341**, 537–40(2013).
[30]X. Cai, J. Wang, M. J. Strain , B. Johnson-Morris, J. Zhu, M. Sorel, J. L. O'Brien, M. G. Thompson, S. Yu, Science **338**, 363–366(2012).
[31]P. Miao, Z. Zhang, J. Sun, W. Walasik, S. Longhi, N. M. Litchinitser, L. Feng, Science **353**, 464(2016).
[32]B. Bahari, A. Ndao, F.Vallini, A. E. Amili,Y. Fainman, B. Kanté, Science **358**, 636–640(2017).